\documentclass[aip,floatfix,reprint,superscriptaddress]{revtex4-2}

\usepackage{amsmath}
\usepackage{amssymb}
\usepackage{amsthm}
\usepackage{bbm}
\usepackage{color}
\usepackage{float}
\usepackage{graphicx}
\usepackage{hyperref}
\usepackage{physics}
\usepackage[T1]{fontenc}
\usepackage{times}
\usepackage{txfonts}
\usepackage[utf8]{inputenc}
\usepackage{xcolor}
\usepackage{multirow}
\usepackage{graphicx}
\usepackage{array}
\usepackage[normalem]{ulem}

\newcommand{\mgraph}[1]{\vcenter{\hbox{\includegraphics[width=10pt]{#1}}}}

\newlength{\gwidth} 
\hypersetup{
    colorlinks=true,linkcolor=blue,citecolor=blue,
    filecolor=blue,urlcolor=blue,breaklinks=true
}

\newcommand{\orcid}[1]{\href{https://orcid.org/#1}
{\includegraphics[width=7pt]{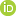}}}

\newtheorem{theorem}{Theorem}

\theoremstyle{definition}
\newtheorem{definition}{Definition}[section]
\newtheorem{proposition}[theorem]{Proposition}

\def\be{\begin{equation}}
\def\ee{\end{equation}}

\def\bc{\begin{center}}
\def\ec{\end{center}}
\def\bal{\begin{align}}
\def\eal{\end{align}}

\begin{document}

\title{Quantum transport in randomized quantum graphs}

\author{Alison A. Silva\orcid{0000-0003-3552-8780}}
\email{alisonantunessilva@gmail.com}
\affiliation{
  Programa de Pós-Graduação em Ciências/Física,
  Universidade Estadual de Ponta Grossa,
  84030-900 Ponta Grossa, Paraná, Brazil
}

\author{D. Bazeia\orcid{0000-0003-1335-3705}}
\email{bazeia@fisica.ufpb.br}
\affiliation{
  Departamento de Física,
  Universidade Federal da Paraíba,
  58051-900 João Pessoa, Paraíba, Brazil
}

\author{Fabiano M. Andrade\orcid{0000-0001-5383-6168}}
\email{fmandrade@uepg.br}
\affiliation{
  Programa de Pós-Graduação em Ciências/Física,
  Universidade Estadual de Ponta Grossa,
  84030-900 Ponta Grossa, Paraná, Brazil
}
\affiliation{
  Departamento de Matemática e Estatística,
  Universidade Estadual de Ponta Grossa,
  84030-900 Ponta Grossa, Paraná, Brazil
}

\date{\today}

\begin{abstract}
This work deals with quantum transport in open quantum graphs.
We consider the case of complete graphs on $n$ vertices with an edge
removed and attached to two leads to represent the entrance and exit
channels, from where we calculate the transmission coefficient.
We include the possibility of several vertices being connected
or not and associate it with a randomization parameter $p$.
To implement the calculation, we had to introduce the transmission
coefficient of randomized quantum graphs (RQG), and we also proposed a
procedure to obtain the exact and approximate but reliable results for
such transmission coefficients.
The main results show that the transport is importantly affected by the
removal of connections between pairs of vertices, but they also indicate
the presence of a region where the transmission is fully suppressed,
even when the number of edge removal is not too small.
\\

\noindent \textbf{APL Quantum 1, 046126 (2024)}; doi: %
\href{https://doi.org/10.1063/5.0239742}
{10.1063/5.0239742}
\end{abstract}

\maketitle

\section{Introduction}
\label{sec:sec1}

The present work aims to investigate how quantum transport in open
quantum graphs is affected by a randomization of edges in a quantum
graph. 
Our motivation is strongly connected with the randomization
model introduced in \cite{PRA.89.052335.2014}, in which randomized graph
states were proposed, and the fact that quantum mechanics provides
interesting perspectives in the field of random networks and graphs
\cite{NP.6.539.2010,TAMS.375.3061.2022},
including the study of transport efficiency in noisy quantum networks
\cite{JPA.56.145301.2023}, and quantum search by quantum walks
\cite{PRL.119.220503.2017,PRL.116.100501.2016}.
It is also inspired by the previous investigations developed in Refs.
\cite{PRA.100.62117.2019,EPJP.135.451.2020}, where transmission
coefficients were obtained for several simple quantum graphs, unveiling
interesting possibilities.
In particular, it was shown that different compositions of simple graphs
may interfere deeply with the output of the transmission coefficient,
including the case where it vanishes, thus contributing to the
construction of filters that block the quantum transport.

The subject is also motivated by the possibility of simulating quantum
graphs experimentally employing microwave networks, as proposed
sometime ago in Ref. \cite{PRE.69.056205.2004}.
This line of investigation and the results obtained in
\cite{EPJP.135.451.2020} by considering simple quantum graphs described
by regular triangles and squares have opened an interesting practical
way of investigating quantum graphs as filters.
The issue was recently considered in
Refs. \cite{EPJP.136.794.2021,PRE.108.034219.2023},
unveiling interesting properties of graphs and networks that can
effectively be regarded for manipulating quantum transport.

The interesting perspectives to consider quantum effects on random
networks and graphs
\cite{NP.6.539.2010,PRA.89.052335.2014,TAMS.375.3061.2022}, together
with the possibility of studying transport efficiency in noisy networks
\cite{JPA.56.145301.2023}, and the results on the use of quantum graphs
as filters \cite{EPJP.135.451.2020}, have motivated us to investigate
another interesting possibility concerning the case of simple graphs in
the presence of imperfections.
This is a hard task because there are
several distinct effects that can be added to the family of
imperfections, including damping, leakage, noise, etc.
Here, however, we shall concentrate on the case of simple quantum
graphs with random edges.
This does not cover the many possibilities, but we hope it will
certainly help us to take a step forward into the complex subject
involving quantum transport in quantum graphs and networks.

To implement the investigation in the present work, we first deal with
the methodology in Sec. \ref{sec:sec2}.
There, we describe the three main issues, the first dealing with graphs,
then with random graphs, and finally, the procedure to calculate the
scattering amplitude in open quantum graphs.
In Sec. \ref{sec:sec3} we introduce the concept of randomized quantum
graphs (RQG), and in Sec. \ref{sec:sec4} we explore this concept with
several results obtained in the present work for the case of complete
graphs on $n$ vertices minus an edge, depicting and highlighting the
most important ones.
We focus on graphs similar to complete graphs because they represent
the most connected simple graphs.
Nonetheless, the methodology can be applied straightforwardly to other
types of graphs.
We finish the investigation in Sec. \ref{sec:sec5}, briefly reviewing the
results obtained and including some possibilities of new directions of
research.

\section{Methodology}
\label{sec:sec2}

The main concern of this work is to calculate the scattering coefficient
of open quantum graphs in the presence of randomized edges.
Methodologically, we first discuss graphs, random graphs, scattering in
open quantum graphs, and then we move on to deal with calculations
involving the presence of random quantum graphs.

\subsection{Graphs}
A graph $G=(V,E)$ is defined as a pair consisting of a finite set of
vertices $V=\{v_1,\ldots,v_n\}$ and a set of edges
$E=\{e_1,\ldots,e_l\}$ \cite{Book.2010.Diestel}.
The order of a graph is the number of vertices denoted by $|V|=n$
and the size of the graph is the number of edges denoted by $|E|=l$.
When a graph $F$ is obtained from $G$ by deleting edges in such
a way that $V_F = V_G$, $F$ is called a spanning subgraph of $G$.
In this latter case, we say that $F$ spans $G$.
Moreover, the topology of the graph is defined by its adjacency
matrix $A(G)$, of dimension $n \times n$, whose elements are given
by $1$ if $i$ is adjacent to $j$ and $0$ otherwise.

\subsection{Random Graphs}
Turning our attention to random graphs, there are two definitions of a
random graph.
One definition is the Erdős-Rényi model \cite{PM.6.290.1959}, in which
there are $\binom{\binom{n}{2}}{l}$ possible random graphs $G(n,l)$
having $n$ vertices and $l$ random edges with the same probability
assigned to each edge.
The second way to define a random graph is given by associating a
probability $p$ to connect each two vertices from $n$ labeled vertices,
and a probability $q=(1-p)$ to not have a connection between them, or
related to the break of the connection.
This latter definition, as proposed by Gilbert \cite{AMS.30.1141.1959},
is called $G(n,p)$ model, is easy to explore \cite{Book.2016.Barabasi}
and will be employed in this work.
The binomial distribution gives the connection between both models
in the form
\begin{equation}
    P_{G} = \tbinom{\binom{n}{2}}{l} p^{l} (1-p)^{\binom{n}{2} - l},
\end{equation}
which gives the probability of obtaining a random graph on $n$ vertices
with exactly $l$ edges \cite{PTRSA.371.20120375.2013}.
It leads to the fact that every graph with exactly $l$ edges has the
same probability to exist;
in this sense, the probability to get a random graph $G(n,p)$ could be
obtained as a weighted sum of the probabilities of graphs obtained by
the $G(n,l)$ randomization \cite{TAMS.375.3061.2022}.

In both models, all the possible edges (given by all the possible pairs
of $n$ vertices) are considered, i.e., going from an empty to a complete
graph on $n$ vertices, $K_n$.
However, we can restrict the possible edges as the elements of the edge
set from a target graph $G$.
This approach allows us to study the removal of edges in this target
graph $G$ by choosing a set of edges $E_{F'} \subseteq E_{G}$ which will
be removed, forming a subgraph $F$ with the same vertices as $G$, but
its edges set is $E_F = E_{G} \setminus E_{F'}$.
In this way, there will be $\tbinom{|E_G|}{|E_{F}|}$ possible random
graphs with exactly $|E_F|$ edges.
This allows us to define the target graph with randomized edge-removal
as a subgraph $F$ of $G$ with the same number of vertices of $G$ but
with $E_F$ edges.
For the Gilbert model, we can also apply this concept of randomization
of a target graph.
Here, we can define the successful connections between the possibles
edges as the set $E_{F}$, while the failed connections as the set
$E_{F'} = E_{G} \setminus E_F$.
Thus,  the probability of obtaining a subgraph $F$ from  $G$ graph is
given by
\begin{equation}
  P_{F} = \tbinom{|E_G|}{|E_{F}|}
  p^{|E_{F}|} (1-p)^{|E_{G} \setminus E_{F}|},
\end{equation}
with $p$ is the probability of the existence of an edge between two
vertices.

\subsection{Scattering in open quantum graphs}
\label{sec:soqg}

Quantum graphs have been studied in connection with Anderson transition
in disordered wires \cite{PRL.48.823.1982,PRL.50.747.1983}, adiabatic
transport \cite{AoP.206.440.1991}, quantum Hall systems
\cite{PRL.79.721.1997}, superlattices \cite{PRB.62.R16294.2000},
quantum wires \cite{JPA.29.87.1996}, and mesoscopic quantum systems
\cite{PRA.30.1982.1984,PRB.42.9009.1990,PRL.92.186801.2004,
  PRB.72.115327.2005,JPA.38.3455.2005}.
The general analysis of the S-matrix was provided in
\cite{JPA.34.10307.2001}.
From a fundamental point of view, quantum graphs have become a
a powerful tool to study different aspects of quantum mechanics.
For instance, to study band spectra properties in networks
\cite{PRL.74.3503.1995}, the relation between periodic orbits and
localization theory \cite{PRL.84.1427.2000}, and chaotic and diffusive
scattering \cite{PRL.85.968.2000,PRE.65.016205.2001}.
The use of quantum graphs as models to study quantum chaos started with
the works of Kottos and Smilansky
\cite{PRL.79.4794.1997,AoP.274.76.1999}.
There, they analyzed the spectral statistics of some simple quantum
graphs, showing that the spectra closely follow the prediction from
the random matrix theory. An interesting result from the
study of quantum graphs is the possibility of obtaining analytical
solutions, even when they present chaotic behavior
\cite{PRL.88.044101.2002,PRE.65.046222.2002,PE.9.523.2001,
PRE.64.036225.2001}.
The spectral analysis of graphs is yet of current interest
\cite{CSF.156.111817.2022}, and quantum graphs are strongly related to
the concept of quantum walks \cite{CP.44.307.2003}, as discussed by
Tanner in Ref. \cite{Incollection.2006.Tanner}.
Since quantum walks are quantum versions of classical random walks, the
case of classical random walks is also of interest.
Moreover, since classical random walks are diffusive, diffusion in
graphs and networks are also of current interest
\cite{CSF.156.111791.2022,PS.99.035120.2024}.
Finally, we can mention that there are some reviews
\cite{JPA.38.R341.2005,AP.55.527.2006,PR.647.1.2016} and a book
\cite{Book.Berkolaiko.2013} approaching different aspects and
properties of quantum graphs.

\begin{figure}[t]
  \centering
  \includegraphics[width=0.8\columnwidth]{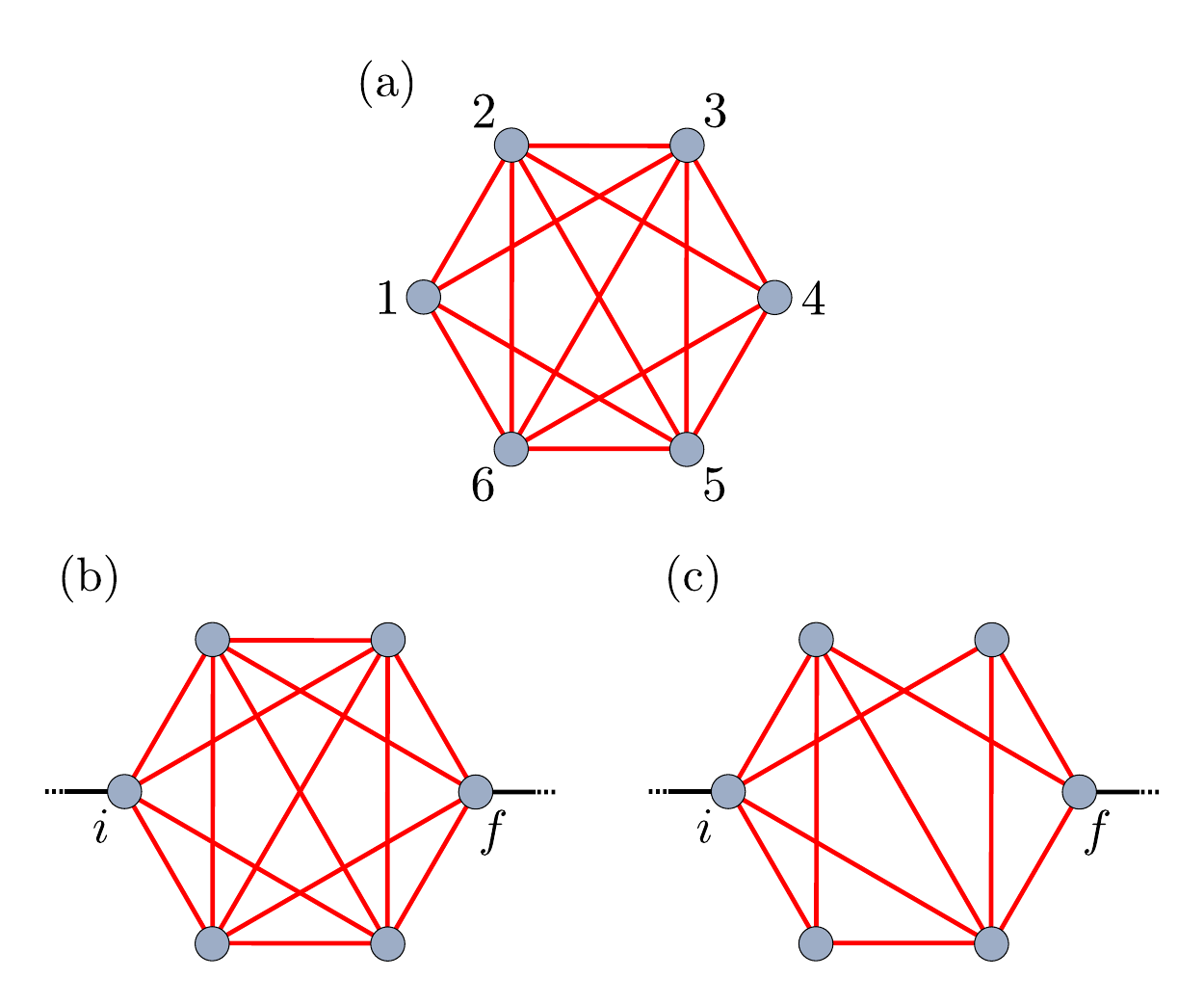}
  \caption{
    (a) Closed quantum graph $\Gamma_{K_6^{e}}$ with $6$ vertices and
    $14$ edges, after removing the edge connecting the vertices $1$ and
    $4$.
    (b) The associated open quantum graph with $2$ leads added, with
    $i=1$ and $f=4$ identify the entrance and exit scattering
    channels, respectively.
    (c) An example of a subgraph with 11 edges.
  }
  \label{fig:fig1}
\end{figure}

To develop our investigation, it is of interest to recall that,
mathematically, a simple quantum graph  $\Gamma_{G}$ consists of
i) a metric graph $\Gamma_{G}$ obtained from a graph $G$ where we
assign positive lengths $\ell_{e_{j}}\in (0,\infty)$ to each edge of the
graph;
ii) a differential operator, $H$; and
iii) a set of boundary conditions (BCs) at the vertices of the quantum
graph, which defines the individual scattering amplitudes at the
vertices \cite{Book.Berkolaiko.2013}.
In this sense, we may say that a quantum graph is a triple
$\{\Gamma_{G},H,\text{BC}\}$.

In this work, we consider the stationary free Schr\"odinger operator
$H=-(\hbar^2/2m)d^2/dx^2$ on each edge, and use
Neumann-Kirchhoff boundary conditions on the corresponding vertices \cite{Incollection.2017.Berkolaiko,Book.Berkolaiko.2013}.
Then, to create an open quantum graph with $c$ scattering channels,
$\Gamma_{G}^c$, which is suitable for studying scattering problems, as
required in the present work, we add $c$ leads (semi-infinite edges) to
its vertices (see Fig. \ref{fig:fig1} for an example with two leads,
$c=2$).
The open quantum graph $\Gamma_G^c$ then represents a scattering system
with $c$ scattering channels, which is characterized by the global energy
dependent scattering matrix $\sigma_{\Gamma_G^c}^{(f,i)}(k)$, where, as
usual, $k=\sqrt{2mE/\hbar^2}$ is the wave number, with $i$ and $f$ being
the entrance and exit scattering channels, respectively.

A suitable method to deal with scattering problems in quantum mechanics
is Green's functions \cite{Book.2006.Economou}.
The Green's function method was applied to quantum graphs in Refs.
\cite{PRE.65.016205.2001,JPA.36.545.2003,PR.647.1.2016,JPA.56.475202.2023}.
In this work, we shall follow the Green's function approach developed in
Ref. \cite{PRA.98.062107.2018} to obtain the global scattering
amplitudes $\sigma_{\Gamma_G^c}^{(f,i)}(k)$.
The method is general and has been successfully used in several studies
of transport in quantum graphs, as in Refs.
\cite{PRA.100.62117.2019,EPJP.135.451.2020,PRA.103.062208.2021,
PE.141.115217.2022}, as well as for modeling the transport in filament
switching \cite{arXiv:2024.06628} and here we introduce a new aspect by
incorporating the randomization of edges.
The exact scattering Green's function for a quantum particle of fixed
wave number $k$, with initial position $x_i$ in the lead $e_i$
and final position $x_f$ in the lead $e_{f}$ can be obtained by using
the adjacency matrix of the underlying graph \cite{PRA.98.062107.2018}.
Thus, the Green's function is written as
\begin{equation}
  G_{\Gamma}(x_f,x_i;k) = \frac{m}{i\hbar^{2} k}
  \sigma_{\Gamma_{G}^{i}}^{(f,i)}(k) e^{i k (x_{i}+x_{f})},
\end{equation}
where
\begin{equation}
\sigma_{\Gamma_{G}^{i}}^{(f,i)}(k) =
\sum_{j \in E_{i}}  A_{ij} p_{ij}(k) t_{i}(k),
\end{equation}
is the global transmission amplitude, with $p_{ij}(k)$ the family
of paths between the vertices $i$ and $j$, which are given by
\begin{equation}
  \label{eq:pij}
  p_{ij}(k)
  = z_{ij} p_{ji}  r_{j}(k)
  +\sum_{l \in {E_{j}^{j,n}}} z_{ij}  A_{jl} p_{jl}  t_{j}(k)
  + z_{ij}  \delta_{jn} t_{n}(k),
\end{equation}
with $z_{ij}= e^{i k \ell_{\{i,j\}}}$ and $\ell_{\{i,j\}}$ the length of
the edge between vertices $i$ and $j$.
The family $p_{ji}(k)$ is given by the same expression above, but
with the swapping of indices $i$ and $j$.
Then, in each vertex $i$ we associated one $p_{ij}(k)$ for every
$j \in E_{i}$, where $E_i$ is the set of adjacent vertices of $i$.
In the above equation, $r_i(k)$ and $t_i(k)$ are the individual
reflection and transmission amplitudes, respectively, associated with
the boundary conditions imposed at the vertex $i$.
For Neumann-Kirchhoff boundary conditions, these individual quantum
amplitudes are independent of $k$, and are explicitly given by
$r_i=({2}/{d_i})-1$ and $t_i={2}/{d_i}$, where $d_i$ is degree of the
vertex $i$.

\section{Randomized Quantum Graphs}
\label{sec:sec3}

\begin{figure}[t]
  \centering
  \includegraphics[width=0.8\columnwidth]{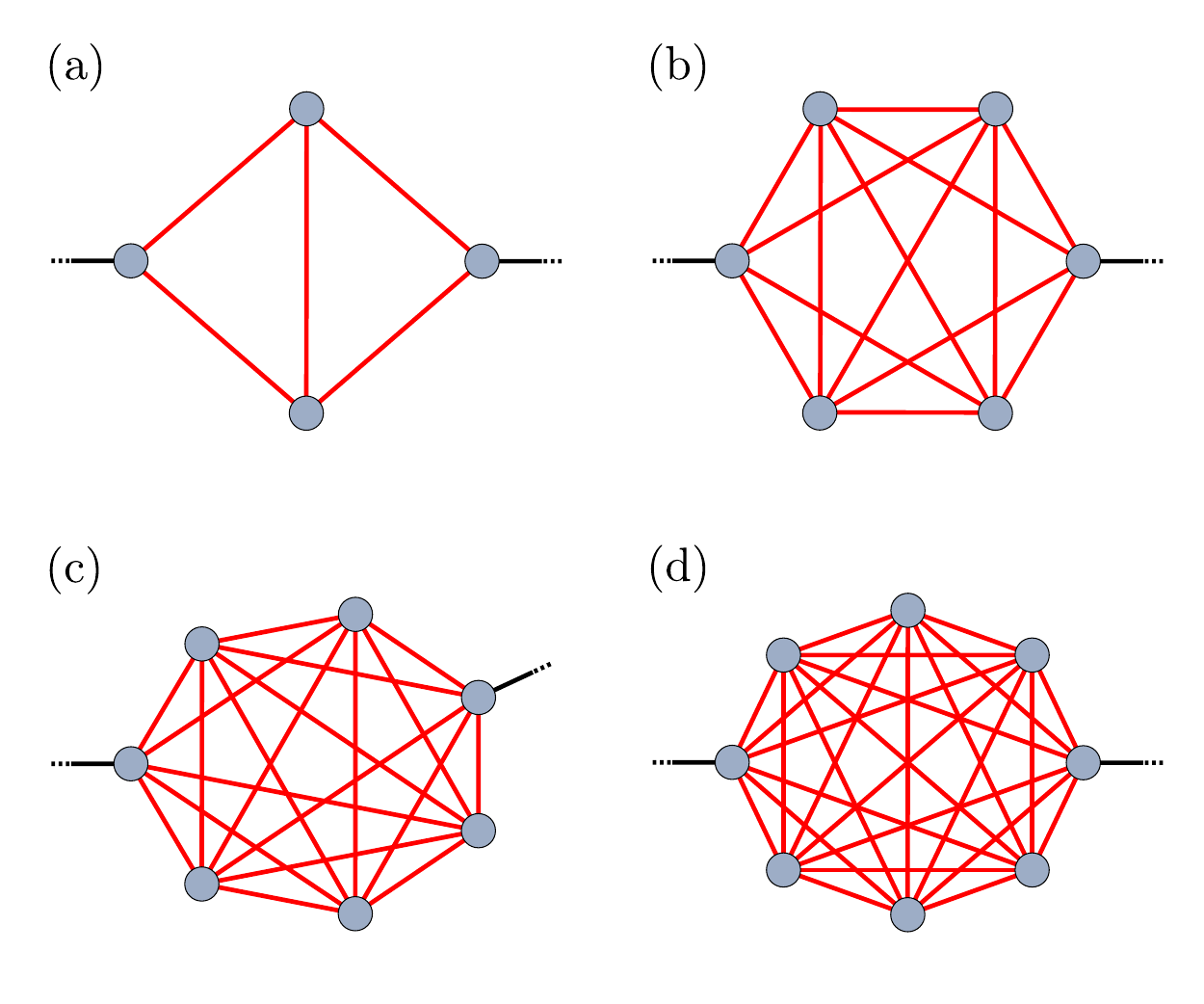}
  \caption{
  The open quantum graphs $\Gamma^2_{K_n^{e}}$ with
    (a) $n=4$ vertices, (b) $n=6$ vertices,
    (c) $n=7$ vertices, and (d) $n=8$ vertices.}
  \label{fig:fig2}
\end{figure}

To apply the concept of random graphs in the model of scattering in
quantum graphs, we can define the randomized quantum graph (RQG) of a
quantum graph $\Gamma_G^c$ by taking all the possible subsets of edges
from $E_{\Gamma_G^c}$, each one defining the set of edges of a quantum
subgraph $\Gamma_F^c$ from $\Gamma_G^c$.
By comparing a quantum subgraph $\Gamma_F^c$ with the larger original
$\Gamma_G^c$, its characteristics as the metrics and BCs from the
original quantum graph ought to be preserved,  but is it possible to
have modifications in the scattering amplitudes at the vertices due to
the removal of edges and the change in the degree of the vertices.
Similarly to graphs, we say that the quantum subgraph $\Gamma_F^c$ spans
the quantum graph $\Gamma_G^c$ when $\Gamma_F^c$ is obtained from
$\Gamma_G^c$ by deleting edges.

We can define the global scattering coefficients of an RQG with random
edge removal by taking all the scattering probabilities for the same
pair of input and output channels in all possible quantum subgraphs,
leading to the following definition.

\begin{definition}
  [Exact transmission coefficient of a RQG]
  \label{def:def1}
  Let $\Gamma_G^c$ be a quantum graph with $c$ scattering channels.
  The exact transmission coefficient associated with the randomized
  quantum graph $R(\Gamma_G^c)$ is defined by
  \begin{equation}
    \label{eq:Tkp}
    T_{R(\Gamma_G^c)}^{(f,i)}(k,p) =
    \sum_{\Gamma_{F}^c \text{ spans } \Gamma_G^c}
    p^{|E_{\Gamma_{F}^c}|} (1-p)^{|E_{\Gamma_G^c} \setminus E_{\Gamma_{F}^c}|}
    \left|\sigma_{\Gamma_{F}^c}^{(f,i)}(k)\right|^2,
  \end{equation}
in which $p$ is the randomization parameter associated with the
probability of the existence of an edge of the original quantum graph.
\end{definition}

In the definition above, the number of quantum subgraphs will depend on
the number of edges present in the original quantum graph.
For instance, in the case of a complete quantum graph on $n$ vertices,
$\Gamma_{K_n}^c$, the number of quantum subgraphs is $2^{\binom{n}{2}}$.
Thus, the number of subgraphs scales exponentially with the number of
vertices and the application of Eq. \eqref{eq:Tkp} can be hard to
implement.
To overcome this issue, here we propose a Monte Carlo method for
calculating the scattering probabilities by taking an ensemble
$S_{|E_{\Gamma_{F}^c}|}$ of size $|S_{|E_{\Gamma_{F}^c}|}|$ of
different quantum subgraphs $\Gamma_{F}^{c}$, each one
having exactly $|E_{\Gamma_{F}^c}|$ random edges,
and calculate the scattering amplitude of each subgraph.
Then, we calculate the average value of these scattering coefficients,
\begin{equation}
\label{eq:sigmaavr}
  \left|\overline{\sigma}_{|E_{\Gamma_F^c}|}^{(f,i)}(k)\right|^{2}
  = \frac{1}{|S_{|E_{\Gamma_{F}^c}|}|}
  \sum_{\Gamma_F^c \in S_{|E_{\Gamma_{F}^c}|}}
  \left|\sigma_{\Gamma_{F}^c}^{(f,i)}(k)\right|^2.
\end{equation}
The subgraphs of the ensemble are chosen uniformly and randomly among
all the possible subgraphs $\Gamma_{F}^{c}$ having exactly
$|E_{\Gamma_{F}^c}|$ edges, without repetition and taking into account
all the isomorphic subgraphs.
We proceed in this manner as determining whether two subgraphs are
isomorphic is a hard problem \cite{CA.63.128.2020}.
Based on this, we propose an approximated method for calculating the
transmission coefficient for RQG.
The proof of the next statement is presented in the Appendix.
\begin{proposition}
  [Approximated transmission coefficient for RQG]
  \label{prop:prop1}
  Let $\Gamma_G^c$ be a quantum graph with $c$ scattering channels and
  $ \mathbb{S}_{|E_{\Gamma_F^c}|}$ the set of all subgraphs $\Gamma_F^c$
  of the quantum graph $\Gamma_G^c$ with a given number of edges
  $|E_{\Gamma_F^c}|$.
  Take an ensemble
  $S_{|E_{\Gamma_F^c}|} \subset \mathbb{S}_{|E_{\Gamma_F^c}|}$, of size
  $|S_{|E_{\Gamma_{F}^c}|}|$ of quantum subgraphs $\Gamma_{F}^c$ with a
  fixed number $|E_{\Gamma_{F}^c}|$ of edges and calculate the average
  scattering coefficient using Eq. \eqref{eq:sigmaavr}.
  The approximated transmission coefficient associated with the
  randomized quantum graph $R(\Gamma_G^c)$ is
  \begin{equation}
  \label{eq:AvrgRandomTransm}
   \mathcal{T}_{R(\Gamma_G^c)}^{(f,i)}(k,p)
  = \sum_{|E_{\Gamma_{F}^{c}}|= 0}^{|E_{\Gamma_{G}^{c}}|}
  \tbinom{|E_{\Gamma_{G}^{c}}|}  {|E_{\Gamma_{F}^{c}}|}
  p^{|E_{\Gamma_{F}^{c}}|}
  (1-p)^{|E_{\Gamma_{G}^{c}} \setminus E_{\Gamma_{F}^{c}}|}
  \left|\overline{\sigma}_{|E_{\Gamma_{F}^{c}}|}^{(f,i)}(k)\right|^2.
\end{equation}
As the ensembles grow larger, $\mathcal{T}_{R(\Gamma_G^c)}^{(f,i)}(k,p)$
converges even closer to $T_{R(\Gamma_G^c)}^{(f,i)}(k,p)$.
\end{proposition}

\section{Results}
\label{sec:sec4}

To study the behavior of the scattering in RQG, we focus on the
family of complete graphs on $n$ vertices,  $K_n$, but with the removal
of the edge $e=\{i,f\}$, as illustrated in Fig. \ref{fig:fig1}(a).
The complete graph on $n$ vertices with the edge $e$ removed is denoted
by $K_n^{e}=K_n\setminus e$.
Moreover, to transform this graph into an open quantum graph with $2$
scattering channels, defined as $\Gamma^2_{K_n^{e}}$, we have to
associate lengths, and here we assign the same length $\ell$ on all
edges.
Also, we use Neumann-Kirchhoff boundary condition on all vertices
and add two leads to the vertices $i$ and $f$, the same vertices where
we removed the connecting edge $e$; see Fig. \ref{fig:fig1}(b).
This procedure ensures that all vertices have the same degree $n-1$,
then the individual reflection and transmission amplitudes are
$r=2/(n-1)-1$ and $t=2/(n-1)$, respectively.
In Fig. \ref{fig:fig2}, we show some examples of the quantum graphs
$\Gamma^2_{K_n^{e}}$ with $n = 4, 6, 7$ and $8$.

\begin{figure}[t]
  \centering
  \includegraphics[width=0.95\columnwidth]{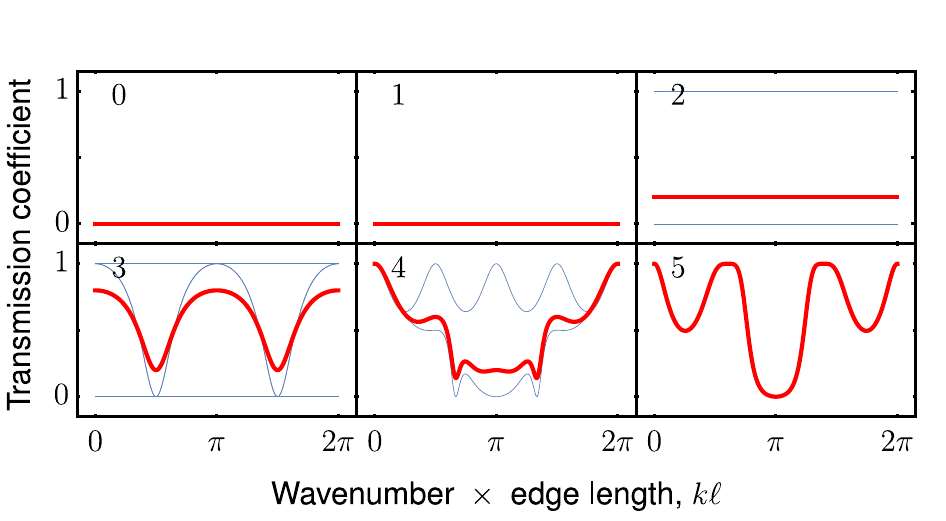}
  \caption{
    Transmission coefficients of quantum subgraphs ensembles with
    $0$ to $5$ edges.
    The blue curves show the individual transmission coefficients
    of the subgraphs with a given number of edges, while the red
    curves represent the average transmission of the ensemble.
  }
  \label{fig:fig3}
\end{figure}

Taking the quantum graph $\Gamma^2_{K_4^e}$ displayed in
Fig. \ref{fig:fig2}(a) as an example and as the number of quantum
subgraphs are not so high, we can calculate the exact transmission for
the randomized quantum graph $R(\Gamma^2_{K_4^e})$ by evaluating the
scattering transmission for all its quantum subgraphs,
Eq. \eqref{eq:Tkp}.
The individual results are shown in Table \ref{tab:tab1}, where we
identified $14$ non-isomorphic quantum subgraphs and evaluated their
number of isomorphisms and transmission amplitudes.
These transmissions can be computed as a function of the number of
edges.
By grouping the graphs with the same number of edges and taking the
average value of the scattering transmission, as defined in
Eq. \eqref{eq:sigmaavr}, we obtain the results presented in
Fig. \ref{fig:fig3}.

\begin{table}
  \caption{
    Non-isomorphic quantum subgraphs from $\Gamma^2_{K_4^{e}}$,
    its number of edges $l$, its number of isomorphic subgraphs (NIS) in
    the ensemble, and the probability associated with the number of
    edges and the corresponding transmission amplitudes. In the
    transmission $z=e^{i k \ell}$.
  }
\begin{center}
\begin{tabular}{ c c c c c}
\hline
\hline
Subgraph & $l$ & NIS  & Probability & Transmission\\
      & &  &             & Amplitude $\sigma_{\Gamma_{F}^2}^{(f,i)}(k)$ \\
  \hline
  \parbox[c]{4em}{\vspace{.15em}\includegraphics[width=.5in]
  {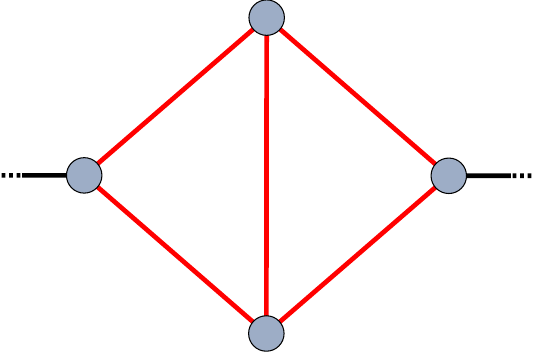}\vspace{.15em}} & $5$ & $1$ & $p^5$ & $\ensuremath{\displaystyle\frac{16(1+z)z^{2}}{27+9z+6z^{2}-6z^{3}-z^{4}-3z^{5}}}$
  \\ \parbox[c]{4em}{\vspace{.15em}\includegraphics[width=.5in]
  {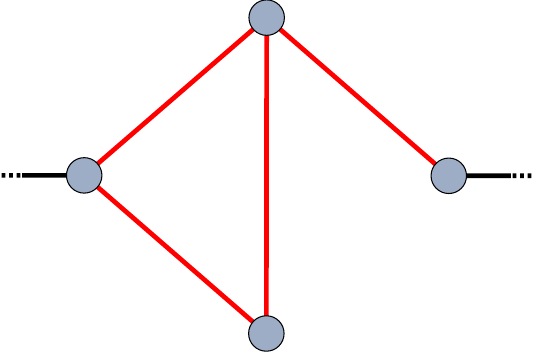}\vspace{.15em}} & $4$ & $4$ & $p^4 (1-p) $ & $\displaystyle\frac{4\left(1+2z+2z^{2}+z^{3}\right)z^{2}}{9+9z+8z^{2}-z^{4}-z^{5}}$
  \\ \parbox[c]{4em}{\vspace{.15em}\includegraphics[width=.5in]
  {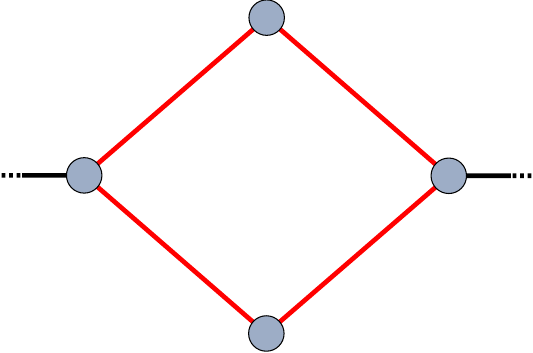}\vspace{.15em}} & $4$ & $1$ & $p^4 (1-p) $ & $\displaystyle\frac{8 z^2}{9-z^4}$
  \\ \parbox[c]{4em}{\vspace{.15em}\includegraphics[width=.5in]
  {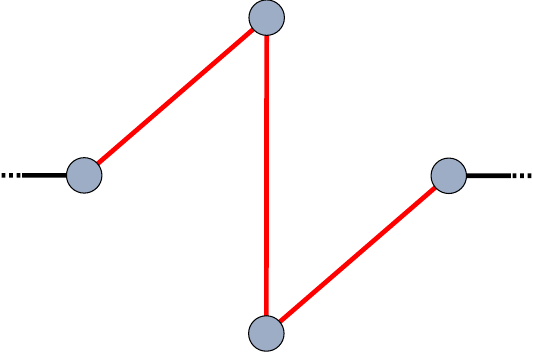}\vspace{.15em}} & $3$ & $2$ & $p^3 (1-p)^2$ & $z^3$
  \\ \parbox[c]{4em}{\vspace{.15em}\includegraphics[width=.5in]
  {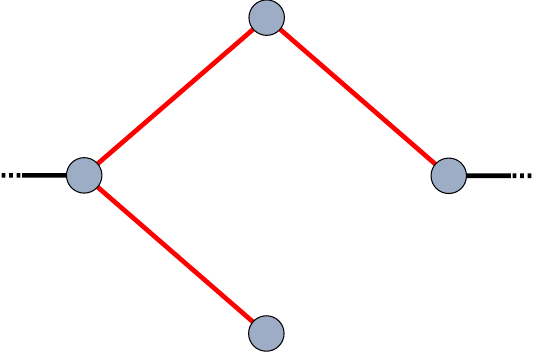}\vspace{.15em}} & $3$ & $4$ & $p^3 (1-p)^2$ & $\displaystyle\frac{2  \left(1+z^2\right)z^2}{3+z^2}$
  \\ \parbox[c]{4em}{\vspace{.15em}\includegraphics[width=.5in]
  {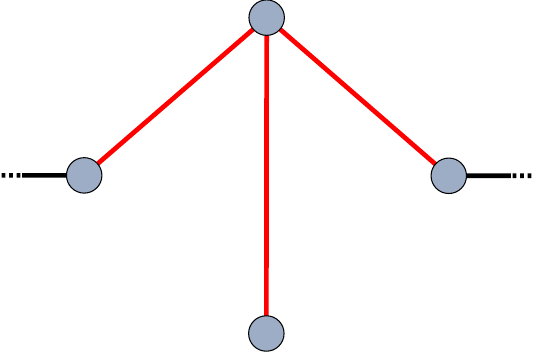}\vspace{.15em}} & $3$ & $2$ & $p^3 (1-p)^2$ & $\displaystyle\frac{2  \left(1+z^2\right)z^2}{3+z^2}$
  \\ \parbox[c]{4em}{\vspace{.15em}\includegraphics[width=.5in]
  {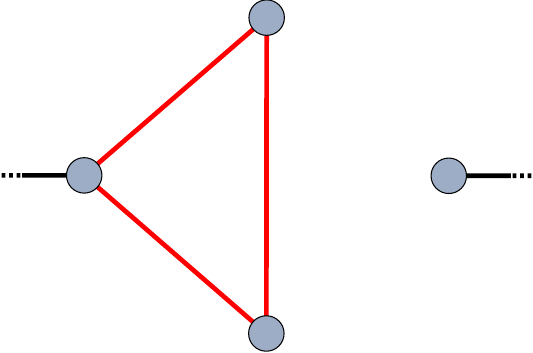}\vspace{.15em}} & $3$ & $2$ & $p^3 (1-p)^2$ & $0$
  \\ \parbox[c]{4em}{\vspace{.15em}\includegraphics[width=.5in]
  {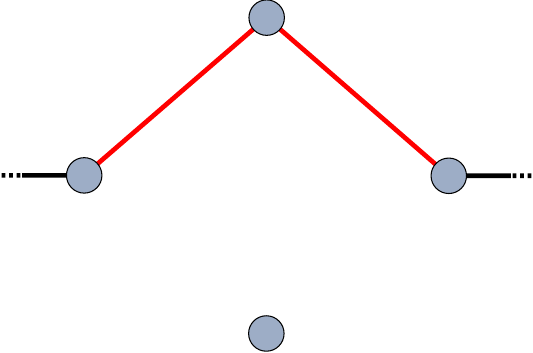}\vspace{.15em}} & $2$ & $2$ & $p^2 (1-p)^3$ & $z^2$
  \\ \parbox[c]{4em}{\vspace{.15em}\includegraphics[width=.5in]
  {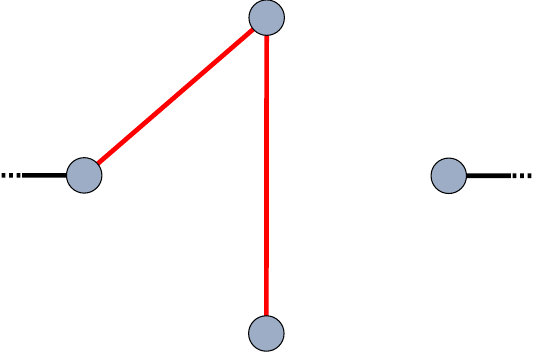}\vspace{.15em}} & $2$ & $4$ & $p^2 (1-p)^3$ & $0$
  \\ \parbox[c]{4em}{\vspace{.15em}\includegraphics[width=.5in]
  {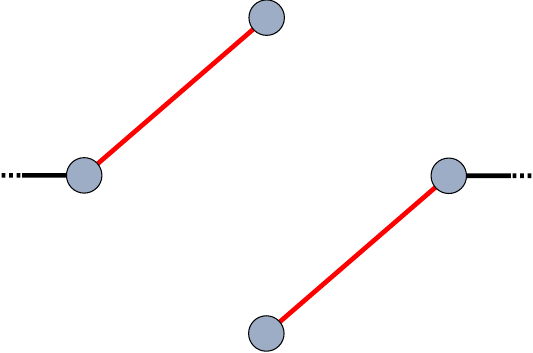}\vspace{.15em}} & $2$ & $2$ & $p^2 (1-p)^3$ & $0$
  \\ \parbox[c]{4em}{\vspace{.15em}\includegraphics[width=.5in]
  {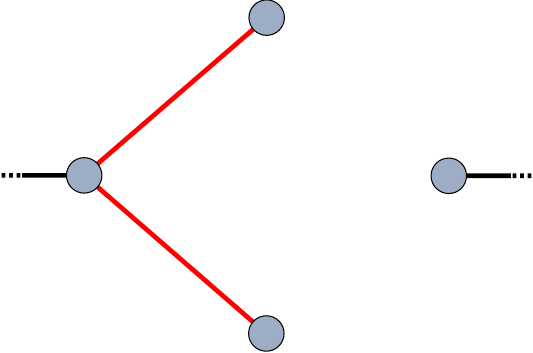}\vspace{.15em}} & $2$ & $2$ & $p^2 (1-p)^3$ & $0$
  \\ \parbox[c]{4em}{\vspace{.15em}\includegraphics[width=.5in]
  {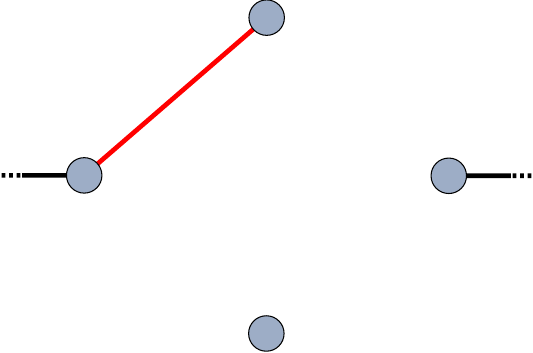}\vspace{.15em}} & $1$ & $4$ & $p (1-p)^4$ & $0$
  \\ \parbox[c]{4em}{\vspace{.15em}\includegraphics[width=.5in]
  {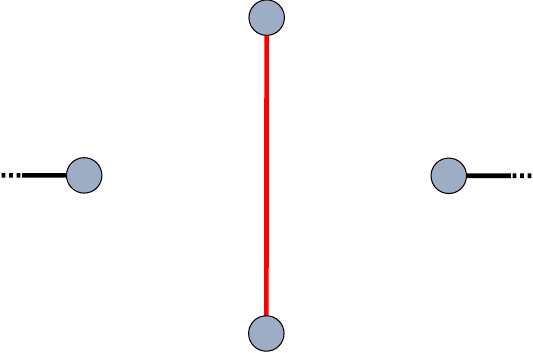}\vspace{.15em}} & $1$ & $1$ & $p (1-p)^4$ & $0$
  \\ \parbox[c]{4em}{\vspace{.15em}\includegraphics[width=.5in]
  {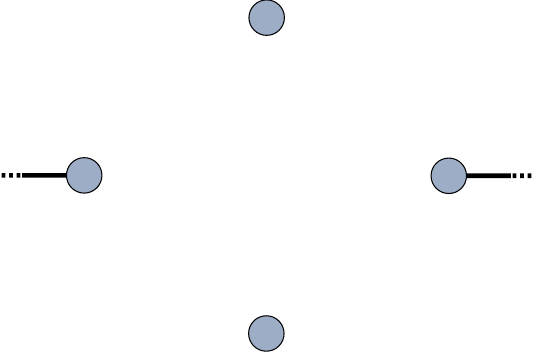}\vspace{.15em}} & $0$ & $1$ & $(1-p)^5$ & $0$
  \\
\hline
\hline
\end{tabular}
  \label{tab:tab1}
\end{center}
\end{table}

\begin{figure}[t]
    \centering
    \includegraphics[width=0.95\columnwidth]{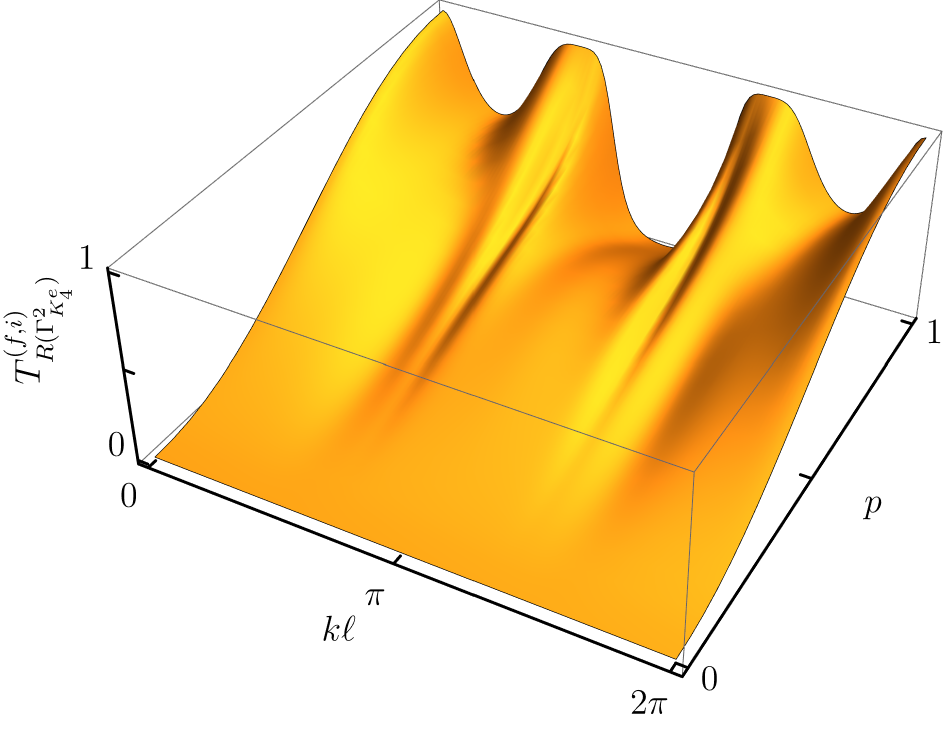}
   \caption{
    The exact transmission coefficient for the RQG $R(\Gamma^2_{K_4^e})$
    as function of the wavenumber $\times$ edge length $k\ell$ and
    randomization parameter $p$.}
    \label{fig:fig4}
\end{figure}

Thus, following Definition \ref{def:def1} the scattering transmission
for the randomized quantum graph $R(\Gamma^2_{K_4^e})$ can be expressed
as the summation of all possible subgraphs weighted by their
probabilities as
\begin{align}
  T_{R(\Gamma_{\mgraph{sg1}}^2)} (k,p) = {}
  &
  p^{5}
    \left|\sigma_{\Gamma_{\mgraph{sg1}}^2}^{(f,i)}(k)\right|^2
    \nonumber\\
  & \quad + p^{4}(1-p)
    \left(4 \left|\sigma_{\Gamma_{\mgraph{sg2}}^2}^{(f,i)}(k)\right|^2
    +
    \left|\sigma_{\Gamma_{\mgraph{sg3}}^2}^{(f,i)}(k)\right|^2
    \right)
    \nonumber\\
  & \quad + p^{3}(1-p)^{2}
    \left(2 \left|\sigma_{\Gamma_{\mgraph{sg4}}^2}^{(f,i)}(k)\right|^2
    +
    4
    \left|\sigma_{\Gamma_{\mgraph{sg5}}^2}^{(f,i)}(k)\right|^2\right)
    \nonumber\\
  & \quad + p^{3}(1-p)^{2}
    \left(2 \left|\sigma_{\Gamma_{\mgraph{sg6}}^2}^{(f,i)}(k)\right|^2
    +
    2\left|\sigma_{\Gamma_{\mgraph{sg7}}^2}^{(f,i)}(k)\right|^2\right) \nonumber\\
  & \quad + p^{2}(1-p)^{3}
    \left(2\left|\sigma_{\Gamma_{\mgraph{sg8}}^2}^{(f,i)}(k)\right|^2
    +
    4\left|\sigma_{\Gamma_{\mgraph{sg9}}^2}^{(f,i)}(k)\right|^2\right)
    \nonumber\\
  & \quad + p^{2}(1-p)^{3}
    \left( 2 \left|\sigma_{\Gamma_{\mgraph{sg10}}^2}^{(f,i)}(k)\right|^2
    +
    2\left|\sigma_{\Gamma_{\mgraph{sg11}}^2}^{(f,i)}(k)\right|^2\right) \nonumber\\
  & \quad + p(1-p)^{4}
    \left( 4 \left|\sigma_{\Gamma_{\mgraph{sg12}}^2}^{(f,i)}(k)\right|^2
    +
    \left|\sigma_{\Gamma_{\mgraph{sg13}}^2}^{(f,i)}(k)\right|^2\right)
    \nonumber\\
  & \quad + (1-p)^{5}
    \left|\sigma_{\Gamma_{\mgraph{sg14}}^2}^{(f,i)}(k)\right|^2.
\end{align}
The corresponding expressions for the transmission amplitudes are presented
in Table \ref{tab:tab1}.
This allows us to evaluate the exact transmission probability for the RQG as
a function of $k\ell$, the wavenumber $\times$ edge length, and the
randomization parameter $p$, as illustrated in Fig. \ref{fig:fig4}.

\begin{figure}[t]
  \centering
  \includegraphics[width=0.95\columnwidth]{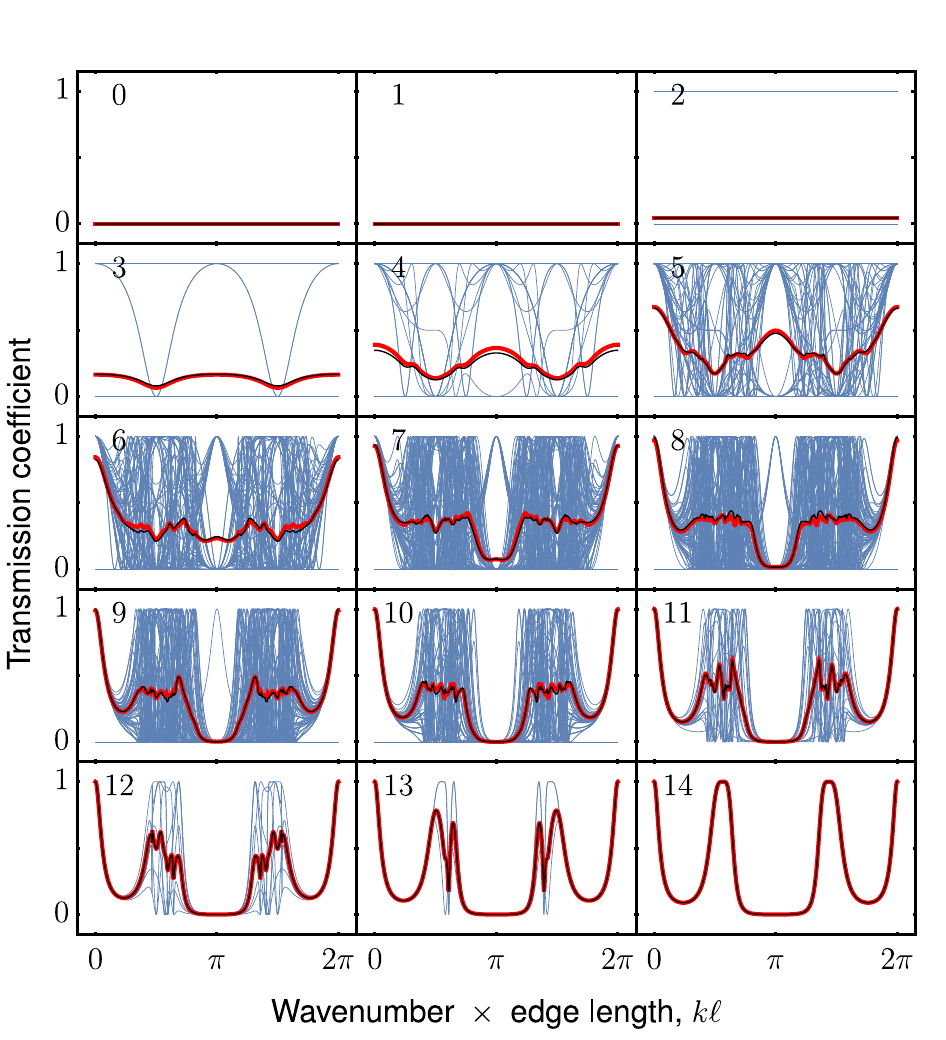}
  \caption{
    Transmission coefficients of ensembles of quantum subgraphs
    $\Gamma_{F}^2$ of the quantum graph $\Gamma_{K_6^e}^2$ with
    $|E_{\Gamma_F^2}|=0, 1, \dots, 14$.
    The blue curves in the background show the transmissions
    coefficients of the corresponding subgraphs, while the red curves
    represent the average transmission of the corresponding ensembles
    and the thin black lines show the average value of
    transmission by taking up to 250 Monte Carlo samples of subgraphs
    from the ensemble.
  }
  \label{fig:fig5}
\end{figure}

\begin{figure}[t]
    \centering
    \includegraphics[width=0.95\columnwidth]{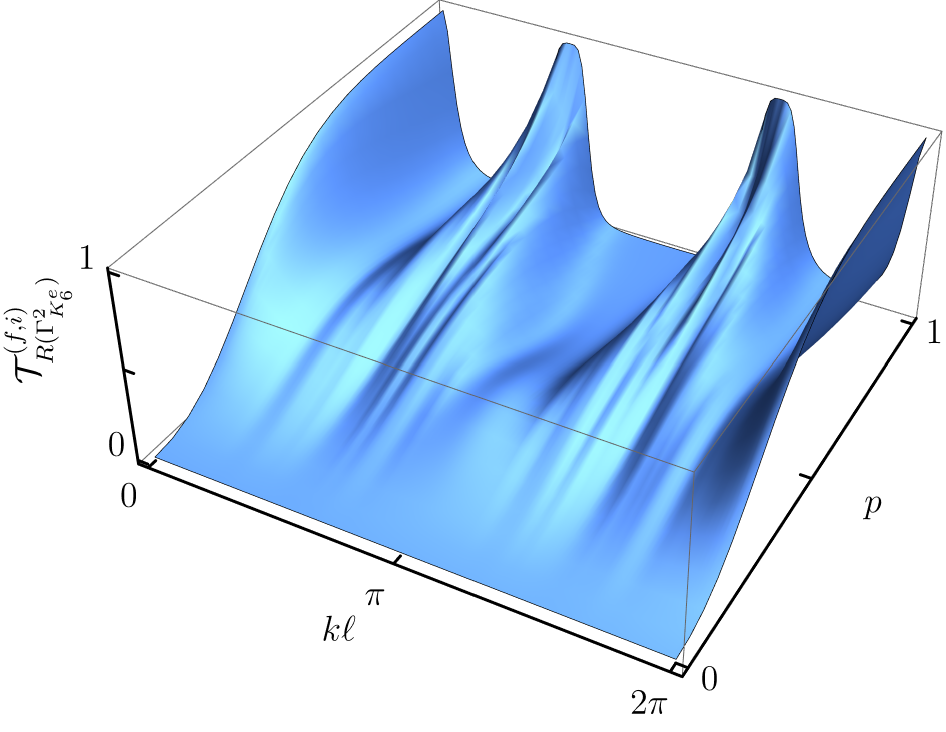}
   \caption{
    Approximated transmission coefficient for the RQG
    $R(\Gamma^2_{K_6^{e}})$ as function of the wavenumber $\times$ edge
    length $k\ell$ and randomization parameter $p$.}
    \label{fig:fig6}
\end{figure}

To compare the exact and the approximated transmission coefficients, now we
consider a more complex case, the complete quantum graph minus an edge with
$n=6$, $K_6^e$.
For $K_6^e$, there are $2^{14} = 16384$ possible subgraphs, which must
be considered for the exact calculation.
In Fig. \ref{fig:fig5}, the background plots (blue curves) represent all
the $16384$ transmission probabilities for the quantum subgraphs of
$\Gamma^2_{K_6^e}$.
The red curves represent the average values of the transmission for
a given number of edges [Eq. \eqref{eq:AvrgRandomTransm}], and the thin
black curves represent the approximate average values of the
transmission by considering ensembles with the same number of edges with
up to $250$ quantum subgraphs.
Finally, the result obtained for the approximated scattering
transmission probability considering these ensembles, which depend on the
number of edges present in the quantum subgraph is shown in
Fig. \ref{fig:fig6}.
Also, we calculated the exact transmission coefficient for this
randomized quantum graph, and the results are almost indistinguishable
from the approximated one, displaying a maximum error of 2.3\%.
This is an indication that our method is adequate.

Some characteristics of these results that can be highlighted are:
i) the transmission is periodic, so we concentrate on the interval
$0\leq k\ell\leq2\pi$;
ii) the transmission vanishes for all $k\ell$ when the number of edges
$|E_{\Gamma_F^2}|$ is less than or equal to $1$, which is expected as
in this case, there is no direct connection between the input and the
output channels, so the graph requires at least two edges to induce
transmission;
iii) for $|E_{\Gamma_F^2}|=2$, due to the Neumann-Kirchhoff
boundary conditions, either the signal is fully transmitted or there is
no transmission at all;
iv) for $k\ell = \pi$, we notice that there is transmission only in the
interval of $2\leq |E_{\Gamma_F^2}| \leq 9$, where the maximum
transmission of the ensemble is reached when $9$ edges are removed;
v) the transmission probability displays a band of full suppression in
the region around $k\ell =\pi$, even when one removes up to $4$
edges due to destructive interference in the transmission.
This is a significant result, informing us about the resilience of the
quantum graph at this region, which allows us to suggest that the graph
can still be used as a filter even in the presence of some noise of
imperfections, i.e., by edge removals.
With these average values of transmission coefficients, we use
Eq. \eqref{eq:AvrgRandomTransm} in Proposition \ref{prop:prop1}.
Working in this direction, we have obtained the results displayed in
Fig. \ref{fig:fig6}.

\begin{figure}[b]
    \centering
    \includegraphics[width=0.95\columnwidth]{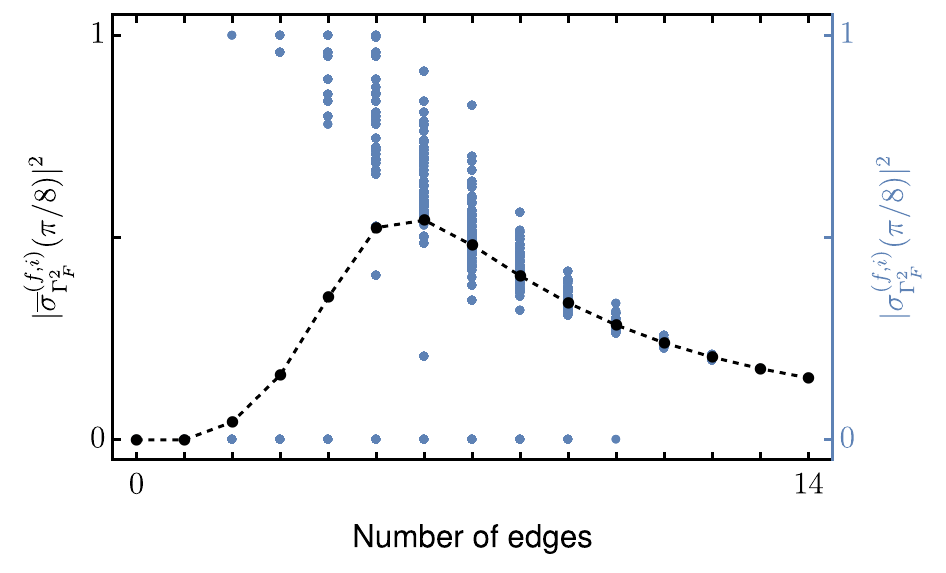}
   \caption{
   The average transmission for the quantum graph $\Gamma^2_{K_6^{e}}$ at
   $k \ell = \pi/8$ as a function of the number of edges}
    \label{fig:fig7}
\end{figure}

\begin{figure}[t]
    \centering
    \includegraphics[width=0.95\columnwidth]{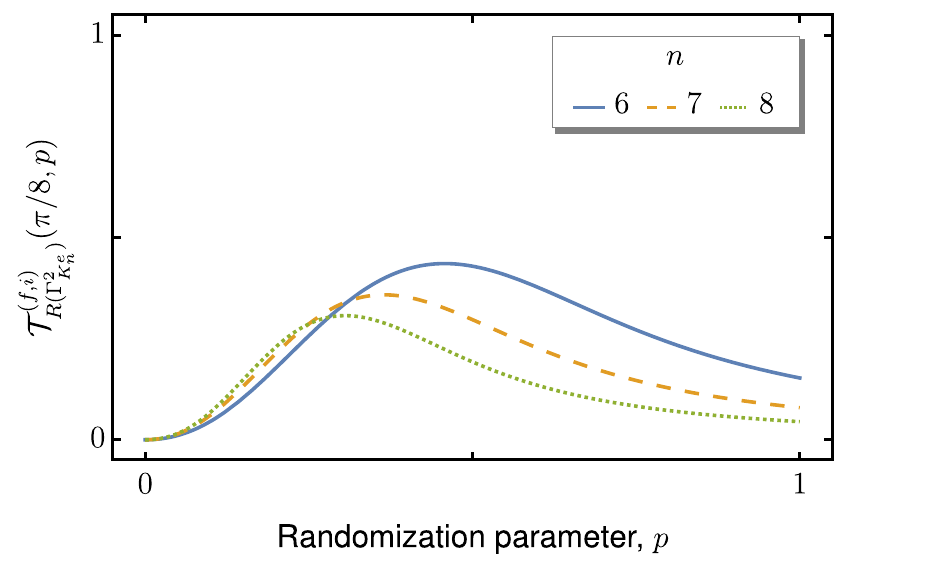}
   \caption{
   Approximated transmission coefficient for the RQGs $R(\Gamma^2_{K_6^{e}})$,
   $R(\Gamma^2_{K_7^{e}})$ and  $R(\Gamma^2_{K_8^{e}})$, for $k \ell =
   \pi/8$, as a function of the randomization parameter$p$.
   }
   \label{fig:fig8}
\end{figure}

Furthermore, we can also evaluate the transmission coefficients for a
given value of $k\ell$ for each one of the cases shown in
Fig. \ref{fig:fig5}, as a function of the number of edges of the graph.
As an example, in Fig. \ref{fig:fig7} we show the values of
the scattering coefficients for $k\ell=\pi/8$ together with the
corresponding average values, depicted with black dots, with the dashed
black line used to guide the eye.
These average values can also be combined using Proposition
\ref{prop:prop1} to allow us to analyze this global behavior as a
function of the probability or randomization parameter $p$ for the
RQG scattering coefficient calculation.
The result is illustrated in Fig. \ref{fig:fig8} for the RQGs
$R(\Gamma^2_{K_6^{e}})$, $R(\Gamma^2_{K_7^{e}})$, and
$R(\Gamma^2_{K_8^{e}})$.

\begin{table}[bh]
    \caption{
      Values of the randomization parameter at the maximum of the approximated transmission
      probability for the RQGs of Figs. \ref{fig:fig8} and \ref{fig:fig10}.
    }
    \begin{center}
      \begin{tabular}{ c c c  c  c c c c}
        \hline
        \hline
        $n$ & $p_{m}$  &  $\mathcal{T}_{R(\Gamma_{K_n^e}^2)}^{(f,i)}(\pi/8,p_m)$
        & \hspace{2cm} &
        $n$ & $p_m$    &  $\mathcal{T}_{R(\Gamma_{K_n^e}^2)}^{(f,i)}(\pi,p_m)$\\
        \hline
        $6$ & $0.458$ &  $0.436$
        & \hspace{1cm} &
        $6$ & $0.334$ &  $0.259$\\
        $7$ & $0.365$ &  $0.358$
        & \hspace{1cm} &
        $7$ & $0.271$ &  $0.239$\\
        $8$ & $0.305$ &  $0.307$
        & \hspace{1cm} &
        $8$ & $0.233$ &  $0.223$\\
        \hline
        \hline
      \end{tabular}
      \label{tab:tab2}
    \end{center}
\end{table}

\begin{figure}[b]
    \centering
    \includegraphics[width=0.95\columnwidth]{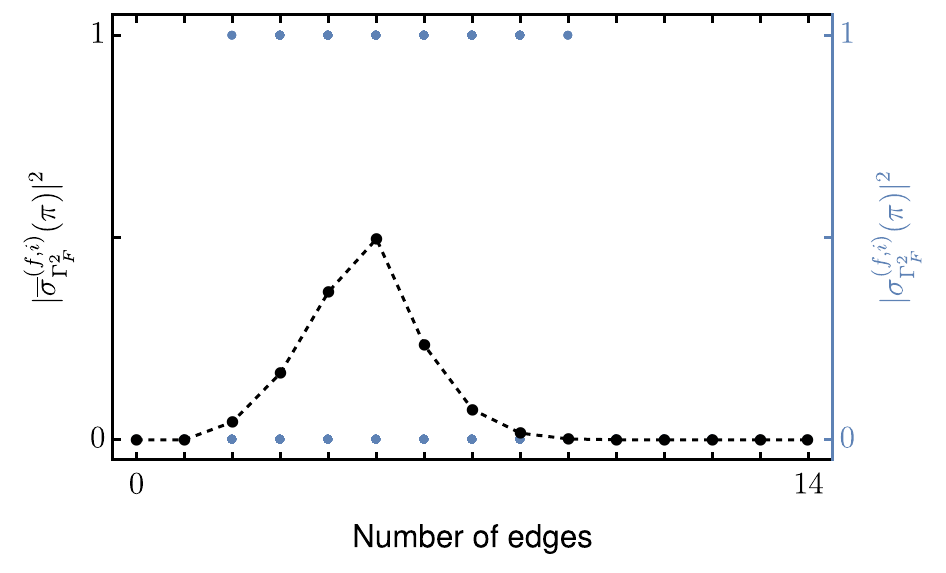}
   \caption{
   The average transmission for the quantum graph $\Gamma^2_{K_6^{e}}$
   at $k \ell = \pi$, as a function of the number of edges.}
    \label{fig:fig9}
\end{figure}

While evaluating the transmission probability for different quantum
subgraphs with different numbers of edges, we noticed that the values
obtained at $k \ell = \pi$ were either $0$ or $1$.
Based on this, we evaluated the distribution of values $0$ and $1$ in
each ensemble at this value of $k \ell$.
In Fig. \ref{fig:fig9} we present the approximated transmission
probability in the ensembles considered for each number of edges.
This behavior is also obtained in other complete quantum graphs minus
an edge with a larger number of vertices, as shown in
Fig. \ref{fig:fig10}, where we considered the RQG
$R(\Gamma^2_{K_6^{e}})$, $R(\Gamma^2_{K_7^{e}})$ and
$R(\Gamma^2_{K_8^{e}})$.
The results show that the maximum value of the transmission is shifted
to the left as we increase the number of vertices.
We can also observe in Figs. \ref{fig:fig8} and \ref{fig:fig10} that
the transmission is not monotonic as a function of $p$, then it is
possible to find the value of the randomization parameter, $p_m$, in
which the transmission is maximum and larger than the transmission for
$p=1.0$.
This shows that RQGs can have larger transmission probabilities when
compared with deterministic quantum graphs (for $p=1.0$).
In Table \ref{tab:tab2} we show the values of $p_m$ and the associated
transmission probabilities for the RQGs of Figs. \ref{fig:fig8} and
\ref{fig:fig10}.
This is an interesting feature that represents an advantage of using
RQGs at some values of $k\ell$.

\section{Ending comments}
\label{sec:sec5}

\begin{figure}[t]
    \centering
    \includegraphics[width=0.95\columnwidth]{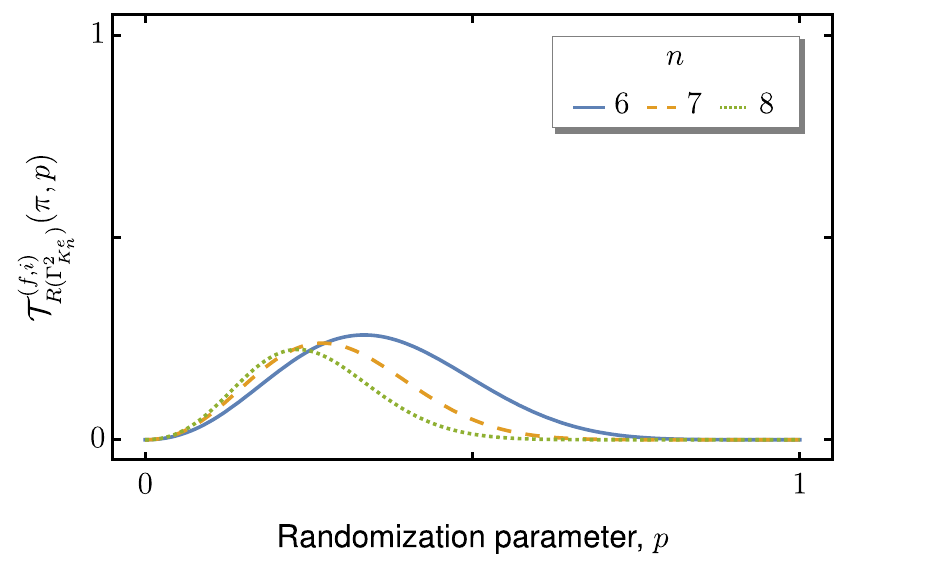}
   \caption{
   Approximated transmission coefficient for  the RQGs $R(\Gamma^2_{K_6^{e}})$,
   $R(\Gamma^2_{K_7^{e}})$ and $R(\Gamma^2_{K_8^{e}})$, for $k \ell =
   \pi$, as a function of the randomization parameter $p$.}
    \label{fig:fig10}
\end{figure}

In this work, we have investigated different aspects of the quantum
transport associated with randomized quantum graphs.
We considered the interesting possibility that random edges in quantum
graphs are taken into account, defining the randomized quantum graphs.
The study is implemented following the methodology developed in
Sec. \ref{sec:sec3}, and the calculations are described under the
procedure illustrated in Fig. \ref{fig:fig1}.
The main results are shown in Figs. \ref{fig:fig5}, \ref{fig:fig6},
\ref{fig:fig7}, \ref{fig:fig8}, \ref{fig:fig9} and \ref{fig:fig10}, with
a specific emphasis is given to the complete graph minus an edge for
$n=6$ vertices.

As the number of subgraphs scales exponentially with the number $n$ of
vertices, we proposed a Monte Carlos method that gives robust results;
this was explicitly shown in the complete quantum graph  minus an edge
with $n=6$ vertices, where the approximate calculation has a maximum
error of 2.3\%, being almost indistinguishable from the exact result.
This indicates that the quantum graphs approached in this work, our
methodology will give even better results when one increases
the number of edges in the quantum graph.
With this motivation, we also studied distinct aspects of the complete
quantum graphs minus an edge $\Gamma^2_{K_7^{e}}$ and $\Gamma^2_{K_8^{e}}$,
as we displayed in Figs. \ref{fig:fig8} and \ref{fig:fig10}.
The results show that the transmission depends on the randomization
parameter $p$, with the corresponding maximum value of the transmission
decreasing as one increases $n$.

We have identified an interesting behavior in the transmission, which is
entirely suppressed for the incoming signal in the region around
$k\ell=\pi$, in the presence of edge removal, even when the removal is not
too small.
This suggests that the use of quantum graphs as filters may possess
robust practical importance, yet in the presence of noise or imperfections,
here considered as edge removal, showing that the system is resilient to
this kind of imperfection in this region.
In Table \ref{tab:tab2} we shown that RQGs can have transmission
probabilities for $0<p<1$ that is higher than deterministic graphs,
which occurs when $p=1$.
It can also motivate research on other types of quantum graphs and the
possibility of using different edge lengths or dressed quantum graphs.
The way the randomization of quantum graphs is defined here also allows
us to randomize a given set of edges of the graph instead of all edges
of the graph, letting us study the introduction of defects in some parts
of the graph in a way similar to the modifications considered in
\cite{PS.98.024005.2023}, e.g., in a given graph, some
edges can be added or removed randomly, and this enables the study of the
impact of the presence of these random edges in the transmission
probability.
These and other generalizations will be reported in future works.

\section*{Acknowledgments}
This work was partially supported by
Coordenação de Aperfeiçoamento de Pessoal de Nível Superior
(CAPES, Finance Code 001).
It was also supported by Conselho Nacional de Desenvolvimento
Científico Tecnológico, Instituto Nacional de Ciência e Tecnologia de
Informação Quântica (INCT-IQ), and Paraiba State Research Foundation
(FAPESQ-PB, Grant 0015/2019).
DB and FMA also acknowledge financial support by CNPq Grants
303469/2019-6 (DB) and 313124/2023-0 (FMA).

\appendix*
\section{Validation of Proposition \ref{prop:prop1}}
\begin{proof}
The Proposition \ref{prop:prop1} can be proofed as follows: Substituting Eq. \eqref{eq:sigmaavr} in Eq. \eqref{eq:AvrgRandomTransm}
the approximated transmission probability can be rewritten as
  \begin{align}
   \mathcal{T}_{R(\Gamma_G^c)}^{(f,i)}(k,p)
  = {} & \sum_{|E_{\Gamma_{F}^{c}}|= 0}^{|E_{\Gamma_{G}^{c}}|}
  \binom{|E_{\Gamma_{G}^{c}}|}  {|E_{\Gamma_{F}^{c}}|}
  p^{|E_{\Gamma_{F}^{c}}|}
  (1-p)^{|E_{\Gamma_{G}^{c}} \setminus E_{\Gamma_{F}^{c}}|}\nonumber\\
  & \times
  \frac{1}{|S_{|E_{\Gamma_{F}^c}|}|}
  \sum_{\Gamma_F^c \in S_{|E_{\Gamma_{F}^c}|}}
  \left|\sigma_{\Gamma_{F}^c}^{(f,i)}(k)\right|^2.
\end{align}
Being $\mathbb{S}_{|E_{\Gamma_{F}^c}|}$ the set of all subgraphs
$\Gamma_{F}^c$ with number of edges $|E_{\Gamma_{F}^c}|$,
when
\begin{equation}
\label{eq:limit}
|S_{|E_{\Gamma_F^c}|}|\to |\mathbb{S}_{|E_{\Gamma_{F}^c}|}|
=\binom{|E_{\Gamma_{G}^{c}}|}{|E_{\Gamma_{F}^{c}}|},
\end{equation}
we have
\begin{equation}
\mathcal{T}_{R(\Gamma_G^c)}^{(f,i)}(k,p)
  = \sum_{|E_{\Gamma_{F}^{c}}|= 0}^{|E_{\Gamma_{G}^{c}}|}
    p^{|E_{\Gamma_{F}^{c}}|}
  (1-p)^{|E_{\Gamma_{G}^{c}} \setminus E_{\Gamma_{F}^{c}}|}
  \sum_{\Gamma_F^c \in \mathbb{S}_{|E_{\Gamma_{F}^c}|}}\left|\sigma_{\Gamma_{F}^c}^{(f,i)}(k)\right|^2.
\end{equation}
The first sum over the number of edges in the quantum subgraphs
$\Gamma_F^c$, and the second one over all the possible quantum subgraphs
with a given number of edges $E_{\Gamma_F^c}$, is equivalent to the sum
over all the spanning subgraphs of $\Gamma_G^c$.
Then, in the limit given by \eqref{eq:limit} we have
\begin{align}
    \mathcal{T}_{R(\Gamma_G^c)}^{(f,i)}(k,p) = {} &
     \sum_{\Gamma_{F}^c \text{ spans } \Gamma_G^c}
    p^{|E_{\Gamma_{F}^c}|} (1-p)^{|E_{\Gamma_G^c} \setminus E_{\Gamma_{F}^c}|}
    \left|\sigma_{\Gamma_{F}^c}^{(f,i)}(k)\right|^2 \nonumber \\
    = {} & T_{R(\Gamma_G^c)}^{(f,i)}(k,p).
\end{align}
\end{proof}

%

\end{document}